\documentclass{appolb}
\usepackage{epsfig}

\newcommand\sfrac[2]{{\textstyle \frac{#1}{#2}}}

\begin{document}
\title{Parametrizing the Neutrino sector%
\thanks{Presented at the XXXV International Conference of 
Theoretical Physics, MATTER TO THE DEEPEST: Recent Developments in 
Physics of Fundamental Interactions, USTRON'11 }%
}
\author{Thomas Gajdosik, Andrius Juodagalvis, \\ Darius Jur\v{c}iukonis, Tomas Sabonis
\address{Vilnius University, Universiteto 3, LT-01513, Vilnius, Lithuania}
}
\maketitle
\begin{abstract}
The original Standard Model has massless neutrinos, but the observation 
of neutrino oscillations requires that neutrinos are massive. The simple 
extension of adding gauge singlet fermions to the particle spectrum 
allows normal Yukawa mass terms for neutrinos. The seesaw mechanism then 
suggests an explanation for the observed smallness of the neutrino masses.
After reviewing the framework of the seesaw we suggest a parametrization 
that directly exhibits the smallness of the mass ratios in the seesaw for 
an arbitrary number of singlet fermions 
and we present our plans to perform calculations for a process that 
might be studied at the LHC.
\end{abstract}
  
\section{Space-time and fermions}
In the last century symmetries became more and more important to describe
nature. Particle physics in particular experienced the need to use 
the symmetry group of Special Relativity from the beginning, as subatomic 
particles travel most of the time at velocities close to the speed of light.
This in turn required that particles are described as representations of the
homogeneous Lorentz group $SO(3,1)$, which is locally isomorphic to the 
product of two rotation groups: $SO(3,1) \sim SU(2)_{L} \otimes SU(2)_{R}$. 
This product representation tells, that apart from the scalar, which does not
transform under Lorentz transformations, the next simple object can be 
a spinor, that transforms under only one of these two rotations groups. These 
spinors are chiral spinors or Weyl-spinors. The more usual Dirac-spinor $\Psi$ 
can be written as the direct sum of two Weyl-spinors of opposite chirality: 
\begin{equation}
  \Psi 
= (\sfrac{1}{2},0) \oplus (0,\sfrac{1}{2})
= \Psi_{L} + \Psi_{R}
\enspace .
\end{equation}

The two discrete symmetries, parity $P$ and time reversal $T$, act on 
space-time and hence on any dependence of the representation of particles, too.
Charge-conjugation, on the other hand, does not act on space-time, 
but only on the particles, i.e.\ on their creation and annihilation operators. 
But for a Lorentz covariant description this charge conjugation should  
be Lorentz covariant, too. Since spinor representations are usually complex, 
charge conjugation understood as a complex conjugation will also influence 
the representation in which fermion fields are defined. Therefore one has
to define a Lorentz covariant conjugation (LCC) 
\begin{equation}
  {\mathcal{C}} \, \Psi  \, {\mathcal{C}}^{-1} =
  \Psi^{c} := \widehat{\Psi} 
= \gamma^{0} \, \mathbf{C} \, ( \Psi )^{*}
= - {\mathbf{C} } \, ( \bar{\Psi} )^{\top}
\label{LCC}
\end{equation}
for spinor fields. It turns 
out, that the LCC acting on chiral fermions will flip the chirality of the 
fermion:
\begin{equation}
  ( \Psi_{L} )^{c} 
= \gamma^{0} \, {\mathbf{C} } \, ( {P_{L}} \Psi )^{*}
= \gamma^{0} {P_{L}} \, {\mathbf{C} } \, ( \Psi )^{*}
= {P_{R}} \, \gamma^{0} \, {\mathbf{C} } \, ( \Psi )^{*}
= {P_{R}} \, \widehat{\Psi}
\enspace .
\label{LCC on chiral spinor}
\end{equation}
These chiral fermions are the building blocks of the Standard Model 
(SM)~\cite{GWS-SM} as can be seen in~\cite{Jegerlehner:1991dq}. 

A Majorana fermion $\Psi_{M}$ is constrained by a reality condition in the 
same way as a real scalar field compared to a complex scalar field: 
\begin{equation}
  \Psi_{M} 
= 
  \eta_{M} \widehat{\Psi}_{M}
=
  \eta_{M} \, \gamma^{0} \, \mathbf{C} \, ( \Psi_{M} )^{*}
\label{Majorana condition}
\end{equation}
with an arbitrary phase $\eta_{M}$. That reduces the four degrees of freedom
of the Dirac fermion to only two degrees of freedom of a Majorana spinor, 
like the two degrees of freedom of a Weyl spinor. But the LCC changes 
the chirality. Therefore a Majorana fermion cannot be a chiral fermion. 
Nevertheless one can define the Majorana fermion by two chiral degrees of 
freedom: 
\begin{equation}
  \Psi_{M} 
= \Psi_{L} + \eta_{M} \widehat{\Psi_{L}} 
= \eta_{M} \widehat{\Psi_{R}} + \Psi_{R}
\enspace .
\label{Majorana from chiral}
\end{equation}
Since a mass term connects both chiralities, a single chiral fermion
cannot support a mass term. But with a Majorana fermion one can write 
a mass term involving only two spinorial degrees of freedom. 
For a didactically extended discussion of 
spinors see~\cite{Pal:2010ih}.

\section{The Standard Model}
The Standard Model (SM)~\cite{GWS-SM} is a chiral quantum gauge field 
theory~\cite{Jegerlehner:1991dq}. 
All fields are massless and obtain mass only through the Higgs 
mechanism~\cite{Higgs-mechanism}. The gauge symmetry 
$SU(3)_{\mathrm{color}} \times SU(2)_{\mathrm{weak}} \times U(1)_{Y}$
is broken by the vacuum expectation value (vev) of the Higgs field to
$SU(3)_{\mathrm{color}} \times U(1)_{\mathrm{em}}$. This leaves the 
gauge bosons of the unbroken gauge symmetries, the gluons and the photon, 
massless. By the choice of the Higgs coupling to the chiral fermions, also 
the quarks and the charged leptons obtain a mass proportional to their 
coupling to the Higgs field. This coupling couples different chiral fermions
and produces pairs of equal mass fermionic states, which group together to 
form the usual Dirac fermions. Since there is no Higgs coupling between 
the lepton doublets and a right chiral $SU(2)_{\mathrm{weak}}$ singlet, 
the neutrinos remain massless in the ''original'' SM~\cite{GWS-SM}.

The SM exhibits an additional global symmetry 
$U(1)_{L} \times U(1)_{R} \sim U(1)_{V} \times U(1)_{A}$. 
The vector combination $U(1)_{V}$ enforces fermion number conservation, 
but the axial vector current $U(1)_{A}$ is anomalously not conserved by 
QCD quantum effects~\cite{Peskin:1995ev}. Since these quantum effects are 
similar to the spontaneous symmetry breaking, one would expect a 
Goldstone like degree of freedom, an axion. Up to now no axion has been 
found~\cite{Baudis:2010zz}.

\subsection{Adding gauge singlet fermions}
One of the major new insights in the last decades is the experimental 
observation of neutrino oscillations~\cite{Fukuda:1998mi}. But the 
massless neutrinos cannot oscillate. 
A very simple extension to the SM is the addition of gauge singlet 
fermions $N$. With these one can write down a mass term for neutrinos 
\begin{equation}
\mathcal{L}_{\mathrm{Yuk},\nu} = 
- \tilde{\phi}^{\dagger} \bar{N} \mathrm{Y}_{\! \nu} \, L_{L} + h.c. 
\enspace ,
\label{nuYukawa}
\end{equation}
in a similar way as for the quarks and the charged leptons 
\begin{equation}
\mathcal{L}_{\mathrm{Yuk}} = 
- {\phi}^{\dagger} \bar{\ell}_{R} \mathrm{Y}_{\! e} \, L_{L} 
- \tilde{\phi}^{\dagger} \bar{u}_{R} \mathrm{Y}_{\! u} \, Q_{L} 
- {\phi}^{\dagger} \bar{d}_{R} \mathrm{Y}_{\! d} \, Q_{L} 
+ h.c.
\enspace ,
\label{Yukawa}
\end{equation}
where $\phi$ is the SM Higgs doublet, $\tilde{\phi} = i\tau_{2}\phi^{*}$; 
$\ell_{R}$, $u_{R}$, and $d_{R}$ are the right handed leptons, up-type and 
down-type quarks, $L_{L}$ and $Q_{L}$ are the left handed lepton and 
quark doublets, and $\mathrm{Y}_{\! k}$ are the respective Yukawa matrices.
Apart from generating the mass 
for neutrinos, which allows for oscillation, this addition does not affect 
other predictions of the overly successful SM\footnote{For the SM being overly 
successful compare the talk by Leszek Roszkowski: ''SUSY in the light of LHC 
and dark matter.''}. Specifically, 
it preserves the global chiral $U(1)_{L} \times U(1)_{R}$ symmetry.

But it leaves open the question, why the neutrino masses are that much smaller
than all the other masses of the SM: after all, the masses of all particles in 
the SM are generated from the single vacuum expectation value $v$ of the 
SM Higgs doublet.

\subsection{Adding a Majorana mass term for the gauge singlet fermions}
Since $N$ is a gauge singlet field and hence electrically neutral, one can 
require a Majorana condition for $N$ and define it with its chiral component:
\begin{equation}
  N 
= N_{L} + \eta_{N} \widehat{N_{L}} 
= \eta_{N} \widehat{N_{R}} + N_{R}
\enspace .
\label{N}
\end{equation}
$N$ being a Majorana fermion, one can also add a Majorana mass term
\begin{equation}
\mathcal{L}_{M} = - \sfrac{1}{2} 
   N_{R}^{\top} \mathbf{C}^{-1} M_{R} N_{R} + h.c. 
\enspace ,
\label{N-mass}
\end{equation}
which breaks the global chiral symmetry explicitly: 
$U(1)_{L} \times U(1)_{R} \to U(1)_{V}$. 
Therefore one no longer would expect an axion. 

This change motivates us to name the model now differently: $\nu$SM.

\section{The seesaw mechanism in the $\nu$SM}
The $n_{R}$ gauge singlets $N$ have the same conserved quantum numbers 
as the three SM neutrinos. Describing the mixing of the neutral fermionic 
fields produces a $(3+n_{R})\times(3+n_{R})$ symmetric mass matrix
\begin{equation} 
  M_{\nu}
=
  \left(\begin{array}{cc}
     M_{L} & M_{D}^{\top} \\
     M_{D} & M_{R}
  \end{array}\right)
\enspace ,
\label{Mnu}
\end{equation}
where $M_{R}$ is the Majorana mass term, eq.(\ref{N-mass}), 
$M_{L} = 0$ at tree level, 
and $M_{D} = v Y_{\nu}$ is the Dirac mass term from the Higgs coupling 
between the lepton doublet and the gauge singlets. In contrast to the 
usual Yukawa matrices for quarks and charged leptons, this Dirac mass term
does not need to be represented with a quadratic matrix.

The most convenient diagonalization of the mass matrix $M_{\nu}$, 
eq.(\ref{Mnu}), for 
arbitrary $n_{R}$ is the Grimus-Lavoura ansatz~\cite{Grimus:2000vj}
\begin{equation} 
  W^{\top} M_{\nu} W
=
  W^{\top} 
  \left(\begin{array}{cc}
     M_{L} & M_{D}^{\top} \\
     M_{D} & M_{R}
  \end{array}\right)
  W
= 
  \left(\begin{array}{cc}
     M_{\ell} & 0 \\
     0 & M_{h}
  \end{array}\right)
\label{Mnu-diagonal}
\end{equation}
with a unitary
\begin{equation} 
  W
= 
  \left(\begin{array}{cc}
     \sqrt{1 - B B^{\dagger} } & B \\
     - B^{\dagger} & \sqrt{1 - B^{\dagger} B }
  \end{array}\right)
\enspace ,
\label{BinW}
\end{equation}
where $B$ is a general complex $3 \times n_{R}$ matrix. 
With the assumption $M_{R} \gg M_{D} \gg M_{L}$ one can expand 
eq.(\ref{Mnu-diagonal}) into a perturbation series and solve the
series recursively for the masses $M_{\ell}$ and $M_{h}$ and the 
mixing matrix $W$, which is completely determined by $B$. 
Seesaw~\cite{seesaw} is the name for the resulting relations 
$M_{h} \approx M_{R}$ and $M_{\ell} \approx M_{D}^{\top} M_{R}^{-1} M_{D}$.

Decomposing $B$ by a singular value decomposition 
\begin{equation} 
B = U_{}^{} S V_{}^{\dagger}
\label{SVD}
\end{equation}
allows us to quantify the parameters of the perturbation expansion
in terms of the singular values $S_{j}$. The lowest order 
of eq.(\ref{Mnu-diagonal}), 
\begin{equation} 
  S V_{}^{\top} M_{h} V_{}^{} S^{\top}
= 
  U_{}^{\top} ( M_{L} - M_{\ell} ) U_{}^{}
\enspace ,
\label{Mnu-lowest-order}
\end{equation}
exhibits the ratio of scales in the seesaw 
\begin{equation} 
  S_{j}^{2} 
= 
  \frac{[ U_{}^{\top} ( M_{L} - M_{\ell} ) U_{}^{} ]_{jj}
       }{[ V_{}^{\top} M_{h} V_{}^{} ]_{jj}
       }
\enspace \sim \enspace \frac{\mathcal{O}(M_{\ell})}{\mathcal{O}(M_{h})}
\enspace \sim \enspace \frac{10^{-9} \, \mathrm{GeV}}{10^{11} \, \mathrm{GeV}}
\enspace \sim \enspace 10^{-20}
\enspace ,
\label{sqrt-singular value}
\end{equation}
expressed by the singular values $S_{j}$.

When comparing $M_{\ell}$ to the oscillation data we see two measured 
differences of squared mass values $\Delta m_{i}^{2}$ for the neutrinos. 
From that we can 
conclude, that at least two values of $M_{\ell}$ have to be non zero. 
At tree level we see, that the number of singular values $S_{j} > 0$ 
gives us the number of non zero mass values in $M_{\ell}$. 
This excludes the possibility of having only a single gauge singlet 
giving masses to the neutrinos at tree level, i.e. $n_{R} = 1$. 
When loop corrections are included, a symmetric $M_{L}$ will be 
generated~\cite{Grimus:1999wm}, which allows more non zero mass values in $M_{\ell}$, 
although the matrix $(M_{L} - M_{\ell})$ still has only rank one and 
hence only a single non zero singular value. 

For two added gauge singlets we can expect two non zero mass values in 
$M_{\ell}$ already at tree level. One neutrino would be expected 
to be massless. The matrices $U$ and $V$, still connected by 
eq.(\ref{Mnu-lowest-order}), allow the parametrization of the 
neutrino sector together with the input of the two masses in $M_{R}$ 
and the two measured differences of squared mass values. Loop corrections
allow to have three non vanishing light neutrino masses. 

In the usually assumed case of $n_{R} = 3$ one can have three 
non zero singular values, giving three non zero mass values in 
$M_{\ell}$ already at tree level. Taking $\Delta m_{i}^{2}$ 
and $M_{h}$ as input parameters 
we still have to choose $V$ and $U$, restricted by 
eq.(\ref{Mnu-lowest-order}), in order to define our model parameters.
A more convenient parametrization, that only works in the case $n_{R} = 3$, 
is the Casas-Ibarra parametrization~\cite{Casas:2001sr}, 
used in~\cite{AristizabalSierra:2011mn}, 
that solves the leading order seesaw equation 
\begin{equation} 
  M_{\ell} 
=
  - M_{D}^{\top} M_{h}^{-1} M_{D} 
\label{nr3seesaw}
\end{equation}
by the ansatz
\begin{equation} 
  M_{D} = i M_{h}^{1/2} \cdot O \cdot M_{\ell}^{1/2}
\label{casas-ibarra}
\end{equation}
with an arbitrary (complex) orthogonal matrix $O$. 
This parametrization is implicitly connected to ours by
\begin{equation} 
  i M_{h}^{1/2} \cdot O \cdot M_{\ell}^{1/2} 
= 
  M_{D}
= 
  M_{h}^{} B^{\dagger} 
= 
  M_{h}^{} V_{}^{} S U_{}^{\dagger} 
\enspace .
\label{connect}
\end{equation}

\section{Neutral fermions in the $\nu$SM}
Since only mass eigenstates describe the physical particles, we have to 
diagonalize the mass matrices of all fields in the $\nu$SM. The chiral 
fields of quarks and charged leptons have mass matrices, that only connect
the left chiral with the right chiral degrees of freedom. This pairing gives 
a symmetric mass matrix with pairs of equal mass values for the left chiral 
and the right chiral mass eigenstates. Therefore one can describe the charged
leptons and the quarks by massive Dirac spinors with four degrees of freedom 
each. 

The $n_{R}$ gauge singlets together with the three neutral leptons form 
also a symmetric mass matrix, but due to the Majorana mass term for the
gauge singlets, the mass eigenvalues do not need to form pairs. Therefore
one gets $(3+n_{R})$ massive fermions with only two degrees of freedom 
each, which can be understood as Majorana fermions. 
There will be three light mass 
values in $M_{\ell}$ and the $n_{R}$ heavy mass values in $M_{h}$. 

The oscillation relevant mixing, described by the neutrino flavor mixing 
matrix, or \emph{Pontecorvo-Maki-Nakagawa-Sakata} matrix 
$U_{\mathrm{PMNS}}$~\cite{PMNS}, 
comes from the coupling of the charged leptonic current 
\begin{equation} 
  W^{-}_{\mu} \bar{\ell}_{L} \gamma^{\mu} P_{L} \nu_{L} 
+
  W^{+}_{\mu} \bar{\nu}_{L} \gamma^{\mu} P_{L} \ell_{L} 
\enspace ,
\label{charged current}
\end{equation}
$\ell_{L} (\nu_{L})$ being the charged (neutral) part of the lepton 
doublet $L_{L}$.
In terms of mass eigenstates, the neutral lepton state $\nu_{L}$ is not 
made up of only the three neutrinos, but has 
also a tiny admixture of the singlet fermions $N$. Inverting the 
mixing of the neutral mass eigenstates 
\begin{equation} 
  \chi 
= \left(\begin{array}{c} \chi_{\mathrm{light}}
                      \\ \chi_{\mathrm{heavy}} \end{array}\right)
= W^{\dagger} \left(\begin{array}{c} \nu_{L} \\ N_{R} \end{array}\right)
\label{mass-flavor-mixing}
\end{equation}
with the parts of eq.(\ref{BinW}) gives
\begin{equation} 
  \nu_{L} 
= 
  \sqrt{1 - B B^{\dagger} } \, P_{L} \, \chi_{\mathrm{light}}
+ B \, P_{L} \, \chi_{\mathrm{heavy}}
\approx
  P_{L} \, \chi_{\mathrm{light}}
\enspace ,
\label{flavor-neutrino}
\end{equation}
which states, that in the basis of the neutral mass eigenstates 
the PMNS matrix is given by the diagonalization matrix of the 
charged leptons
\begin{equation} 
  v Y_{e} 
= 
  U_{eR}^{} \cdot \mathrm{diag}( m_{e} ) \cdot U_{eL}^{\dagger}
= 
  U_{eR}^{} \cdot \mathrm{diag}( m_{e} ) \cdot U_{\mathrm{PMNS}}^{\dagger}
\enspace .
\label{charged-lepton-yukawa}
\end{equation}
So $U_{\mathrm{PMNS}}$ has its origin, like the CKM matrix, in the 
non-alignment of the charged lepton and neutrino mass matrices.

\section{Outlook}
When comparing precision measurements to the predictions of the SM, one has
to include loop corrections, as exemplified in~\cite{Jegerlehner:1991dq}. 
The conceptually simplest renormalization scheme is the on-shell prescription, 
where all external particles are physically measured. These measurements define 
the scale, where the counter terms for the quantum corrections can be calculated. 
For confined quarks this prescription cannot be used, since they cannot be 
observed as free particles, but for the neutrinos the on-shell prescription 
should work very well. 

The large difference in scales encountered in the neutrino sector suggests, 
that the Born approximation, i.e. using only tree level processes, should be
sufficient for the calculations. But as shown in~\cite{Grimus:1999wm}, 
though $M_{L}=0$ at tree level, $\delta M_{L}$ will receive contributions by 
loops with a neutral fermion and a Higgs- or $Z$-boson; these were calculated in
a general framework by Grimus and Lavoura~\cite{Grimus:2002nk}. 
Since these contributions can 
lift the zero mass degeneracy even in the case of $n_{R}=1$ they 
change the picture obtained in the Born approximation completely. The 
Majorana mass term generated in this way is of the size $( v Y_{e} )^{2} / M_{h}$, 
which is the same size as the seesaw generated $M_{\ell}$, and 
it can be included as an effective mass term in $M_{\nu}$. The 
diagonalization with the Grimus-Lavoura ansatz is not changed by these 
quantum corrections. 

We want to generalize the analysis of \cite{AristizabalSierra:2011mn} to include 
the cases for $n_{R} = 1$ or $2$. Specifically we want to look at the process 
\begin{equation} 
  W^{\pm} 
\to 
  \tau^{\pm} + \nu  
\to
  h_{1}^{\pm} + h_{2}^{\mp} + h_{3}^{\pm} + \nu + \nu
\label{W-to-tau-decay}
\end{equation}
and study the $\tau$ polarization coming from the decay of a $W$ 
at the LHC.


\end{document}